\documentclass[11pt]{article}

\usepackage[utf8]{inputenc}
\usepackage[T1]{fontenc}
\usepackage{lmodern}
\usepackage{microtype}

\usepackage[a4paper,margin=1in]{geometry}

\usepackage{amsmath}
\usepackage{amssymb}
\usepackage{textcomp}
\usepackage{siunitx}
\usepackage{graphicx}
\usepackage{caption}
\usepackage{subcaption}
\usepackage{booktabs}
\usepackage{array}

\usepackage{authblk}

\usepackage{cite}

\usepackage{xcolor}
\definecolor{linkblue}{RGB}{0,71,171}
\usepackage[
  colorlinks = true,
  linkcolor  = linkblue,
  citecolor  = linkblue,
  urlcolor   = linkblue,
]{hyperref}

\title{Radiometric Thermal Imaging Dataset of Laboratory Rats\\[2pt]
       with Anatomical Segmentation Masks}

\author[1]{Dima Bykhovsky}
\author[2]{Evyatar Chaimoff} 
\author[2]{Pe'er Eden} 
 \author[2]{Tom Simkin}  
\author[2]{Oshrit Hoffer}
\author[4]{Shahar Cohen}
\author[3,4]{Bar Eilat Yogev} 
\author[3,4]{Gal Levi}
\author[4]{Noa Efroni} 
\author[4,5]{Doron Todder}
\author[3,4,5]{Hagit Cohen} 

\affil[1]{Department of Electrical and Electronics Engineering, Shamoon College of Engineering, Israel}
\affil[2]{School of Electrical Engineering, Afeka College of Engineering, Israel}
\affil[3]{Department of Psychology, Ben-Gurion University of the Negev, Beer-Sheva, Israel}
\affil[4]{Anxiety and Stress Research Unit, Ministry of Health, Be'erot Mental Health Center, Beer-Sheva, Israel}
\affil[5]{Faculty of Health Sciences, Ben-Gurion University of the Negev, Beer-Sheva, Israel} 

\hypersetup{
  pdftitle  = {A Radiometric Thermal Imaging Dataset of Laboratory Rats with Anatomical Segmentation Masks},
  pdfauthor = {Dima Bykhovsky, Evyatar Chaimoff, Oshrit Hoffer},
}

\date{}

\begin{document}

\maketitle

\section*{Abstract}
Infrared thermography provides a contact-free, restraint-free method to record surface temperatures. It serves as a valuable marker for thermoregulatory responses in laboratory animal stress and pharmacology research. However, the analysis of these images is currently bottlenecked by the manual delineation of anatomical regions. To date, no public dataset has provided paired radiometric thermal frames of rats with pixel-level body-part labels.
We present a dataset of 1,655 quality-controlled radiometric thermal frames from 25 laboratory rats. Each frame is paired with a dense four-class anatomical segmentation mask (background, head, body, and tail) and the raw $480 \times 640$ temperature matrix (rows $\times$ columns) in degrees Celsius. 
This ensures every label is registered directly to the physical temperature it describes rather than a color-mapped rendering.
The frames originate from two pharmacological cohorts where interventions alter thermoregulation in opposite directions: ethanol, which induces peripheral vasodilation, and ketamine, which affects central thermoregulation. This provides a wide and physiologically diverse range of surface temperature regimes.
Aggregated across the dataset, the per-class temperatures follow a head~$>$~body~$>$~tail ordering in physical units. To demonstrate that the data support pixel-level segmentation directly from the radiometric channel, we present an exploratory U-Net segmentation pipeline that attains a subject-level cross-validated mean intersection-over-union of $0.895 \pm 0.006$. The dataset provides a reuse-ready benchmark for thermal semantic segmentation and for downstream physiological and stress-phenotyping analyses.

\section{Background \& Summary}
Infrared thermography (IRT) is a simple method for monitoring laboratory animals because it is contact-free, silent, and can be performed without restraint or anesthesia. It records surface temperature as a non-invasive readout of the central and peripheral thermoregulatory mechanisms, avoiding confounding factors common in traditional temperature measurements, such as stress-induced hyperthermia~\cite{blenkus2022stress}. For example, this characteristic makes IRT particularly valuable for stress research \cite{yogev2026acute}. Moreover, temperature-related signatures have been linked to various aspects in rats~\cite{vianna2005changes,wongsaengchan2023body,lecorps2016assessment}. 

While IRT acquisition is non-invasive and highly scalable, the analysis phase remains a severe bottleneck. Research protocols generate hundreds of images per animal. However, isolating anatomical regions of interest currently relies on manual labeling that does not scale and introduces rater variability. Prompting recent reviews to call for standardized, automated IRT pipelines~\cite{verduzco2025thermal}.

Automated semantic segmentation could replace manual annotation. However, models pre-trained exclusively on visible-light imagery transfer poorly to the radiometric thermal domain. While deep instance segmentation has been applied to rodents in thermal images, prior work explicitly cites the need for larger and more diverse thermal databases as a principal limitation~\cite{mazurmilecka2020deep}. To date, no public dataset of radiometric thermal frames of rat pairs with multi-class body-part masks exists. Existing thermal animal datasets are either too small, lack pixel-level labels, or are not released publicly.

This data descriptor introduces the public dataset that pairs radiometric thermal images of rats with dense, multi-class anatomical segmentation masks. We provide $1{,}655$ quality-controlled radiometric thermal frames from $25$ rats, each coupled with a dense four-class anatomical mask (background, head, body, and tail) and the raw $480 \times 640$ temperature matrix in degrees Celsius, ensuring every label is registered directly to the physical temperature rather than a color-mapped rendering. The corpus draws on two pharmacological cohorts whose interventions perturb thermoregulation in opposite directions (Section~\ref{subsec:animal-origin}), and samples protocol-timed physiological states (baseline, peri-stimulus, and treatment) across both a prevention and a therapeutic paradigm \cite{yogev2026acute}. As a worked example of the analysis the data afford, we include an exploratory U-Net segmentation pipeline (Section~\ref{sec-tech-validation}).

\section{Methods}
\subsection{Animals}
Male Sprague–Dawley rats (150–200 g, Envigo, Israel) were pairhoused under a 12:12 h light/dark cycle (lights on 07:00) with ad libitum access to
food and water.

\subsection{Experimental background}\label{subsec:animal-origin}
The 25 rats analyzed in this work originated from two independent pharmacological studies conducted at Ben-Gurion University of the Negev: one investigating ketamine and the other investigating ethanol. In both studies, infrared thermal imaging was added as a non-invasive secondary measurement to capture systemic surface-temperature responses to the administered compound, without altering the underlying experimental procedures \cite{yogev2026acute, levy2026impact}. The two interventions produce contrasting thermal effects: ketamine alters central thermoregulation, whereas ethanol induces peripheral vasodilation and a characteristic drop in core temperature~\cite{dominguezoliva2023thermal}. Pooling subjects across these two cohorts is therefore the reason the resulting corpus spans a wide and physiologically diverse range of surface-temperature regimes. 

\subsection{Image acquisition and annotation}
\subsubsection{Imaging}
Acquisitions were performed with a FLIR ONE PRO smartphone-mounted thermal camera, a long-wave infrared (LWIR) uncooled microbolometer that exposes a raw radiometric temperature matrix at a native sensor resolution of $640\times 480$ (width~$\times$~height). The device also offers a Multi-Spectral Dynamic Imaging (MSX) overlay that fuses radiometric infrared with a co-mounted visible-light camera; because the visible-to-infrared parallax of MSX is subject-distance dependent, we bypass MSX for modeling and retain the raw radiometric matrix as the authoritative measurement, while the MSX-augmented JPEGs are kept only as a visual reference for human annotators.

Ambient laboratory temperature was held within $22$-$25\,^{\circ}$C across sessions. A total of $41$ imaging sessions were carried out across the two pharmacological cohorts. Within each session, animals were imaged at protocol-defined timepoints (baseline, peri-stimulus, and treatment phase), so that the released corpus samples physiologically distinct thermal states rather than redundant frames from a single equilibrium.

\subsubsection{Annotation procedure}
Pixel-level masks were drawn by human annotators working from the Jet-colormap PNG rendered from the raw radiometric matrix, not from the MSX JPEG, so that every label is registered to the radiometric pixel it describes. Annotation was performed in the MATLAB Medical Image Labeler application, with four mutually exclusive classes: $0$ (Background), $1$ (Head, including ears), $2$ (Body, comprising back, abdomen, and limbs), and $3$ (Tail). Two quality-control passes were applied before release. First, a triplet-consistency check retained only frames for which a matching thermal rendering, a same-shape four-class mask, and a raw temperature CSV all existed. Second, a model-assisted anomaly review used the segmentation model described in Section~\ref{sec-tech-validation} to flag candidate label errors for human re-inspection, and the corrected masks form the released ground truth.

\section{Data Records}

The released corpus is organized as one top-level subdirectory per animal, named \texttt{Rat1} through \texttt{Rat25}. Each per-rat directory contains three parallel folders, \texttt{CSV/}, \texttt{Mask/}, and \texttt{Thermal Imaging/}, whose contents are paired by a common integer frame index $N$ on a shared $480 \times 640$ pixel grid ($480$ rows $\times$ $640$ columns). The on-disk format and loading conventions for each of the three artifacts are described below.

\subsection{Temperature matrix}
The radiometric measurement for frame $N$ is stored at \texttt{CSV/Thermal\_$N$\_CSV.csv} as an approximately $2.1$\,MB CSV file holding raw temperatures in degrees Celsius on a $480 \times 640$ float grid. The file requires header-aware parsing: line $1$ records the originating file path, line $2$ is empty, and the following $480$ data rows each begin with a row-label column (\texttt{"Frame 1,"} on the first data row and \texttt{","} on the remaining rows) that must be dropped at load time, leaving $640$ floating-point temperature columns.

\subsection{Segmentation mask}
The pixel-level annotation for frame $N$ is stored at \texttt{Mask/Thermal\_$N$.png} as a single-channel \texttt{uint8} PNG encoding a four-class semantic label map with pixel values $\{0,1,2,3\}$ for Background, Head, Body, and Tail respectively. The mask must be read with raw pixel-value preservation, since standard image-loading paths may apply automatic colormap conversion that would corrupt the discrete class labels.

\subsection{Thermal rendering}
The visualization of frame $N$ is stored at \texttt{Thermal Imaging/Thermal\_$N$.png} as an approximately $74$\,kB \texttt{uint8} RGB PNG (JPEG for a subset of animals) obtained by mapping the temperature matrix through the Jet colormap. This rendering served as the visual reference shown to human annotators and is retained for inspection only; analysis code should consume the radiometric CSV rather than this rendering.

\section{Data Overview}\label{sec:data-overview}

\begin{figure}[!h]
    \centering
    \includegraphics[width=0.8\linewidth]{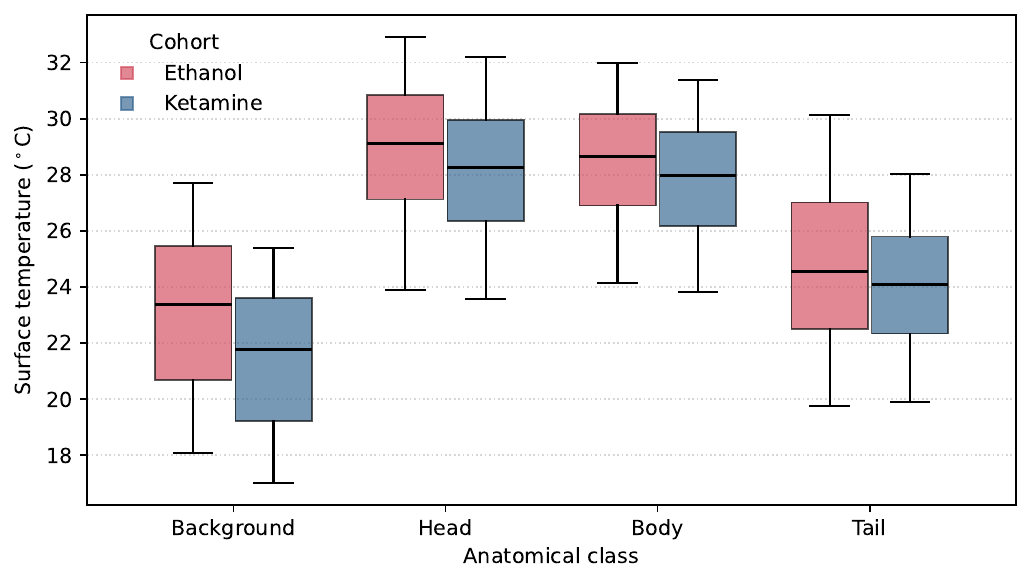}
    \caption{Per-class surface-temperature distribution across the released corpus, split by cohort. Each box spans the inter-quartile range with the median marked, and the whiskers span the 5th--95th percentiles, summarizing every labeled pixel in the dataset. The head~$>$~body~$>$~tail ordering is evident, and the ethanol cohort sits warmer than the ketamine cohort across all classes.}
    \label{fig:class-temp-dist}
\end{figure}

\begin{figure}[!h]
    \centering
    \begin{subfigure}[t]{0.4\linewidth}
        \centering
        \includegraphics[width=\linewidth]{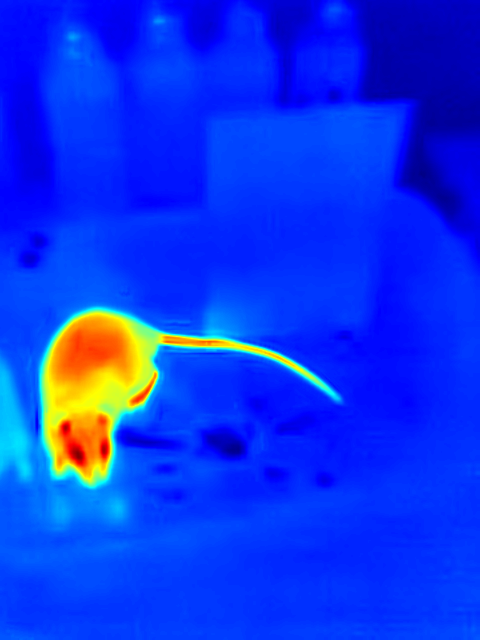}
        \caption{Thermal image (Jet colormap)}
    \end{subfigure}
    \hfill
    \begin{subfigure}[t]{0.4\linewidth}
        \centering
        \includegraphics[width=\linewidth]{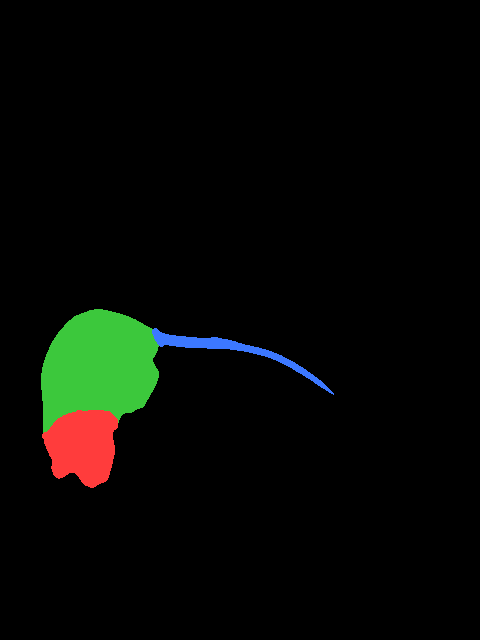}
        \caption{Ground-truth mask}
    \end{subfigure}

    \vspace{0.5em}

    \begin{subfigure}[t]{0.4\linewidth}
        \centering
        \includegraphics[width=\linewidth]{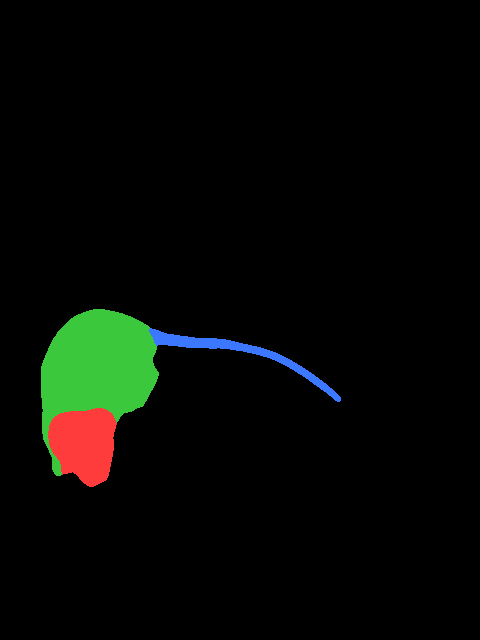}
        \caption{Predicted mask}
    \end{subfigure}
    \hfill
    \begin{subfigure}[t]{0.4\linewidth}
        \centering
        \includegraphics[width=\linewidth]{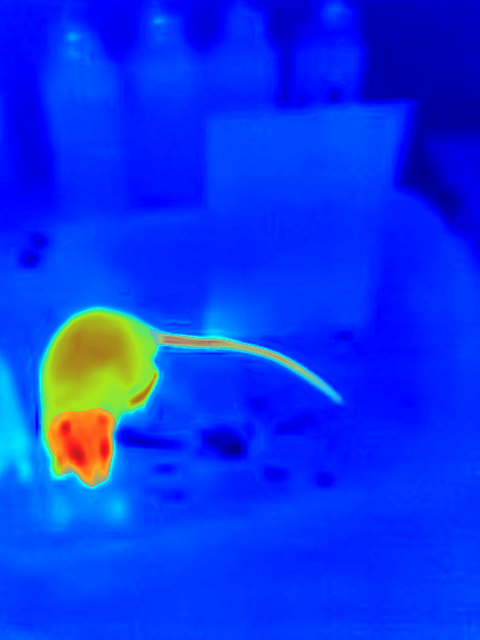}
        \caption{Prediction overlay on thermal}
    \end{subfigure}
    \caption{Example frame from Rat11 (ethanol cohort, frame~1). (a) Thermal image rendered from the radiometric CSV with the Jet colormap (cool $\rightarrow$ warm). (b) Ground-truth semantic segmentation mask with values $0$ (Background), $1$ (Head), $2$ (Body), and $3$ (Tail). (c) Predicted mask from the U-Net pipeline described in Section~\ref{sec-tech-validation}. (d) The predicted mask superimposed on the thermal rendering.}
    \label{fig:dataset-example}
\end{figure}

The corpus consists of infrared thermal recordings of 25 laboratory rats acquired across multiple imaging sessions, for a total of $1{,}655$ quality-controlled frames. Animals originated from the two pharmacological studies as described in Section~\ref{subsec:animal-origin}; imaging timepoints were chosen so that the resulting collection spans diverse surface-temperature regimes across subjects and conditions. The corpus comprises an ethanol cohort of $15$ subjects (Rats 1--15, $307$ frames) and a ketamine cohort of $10$ subjects (Rats 16--25, $1{,}348$ frames), corresponding to the two approvals listed in Section~\ref{subsec:animal-origin}; per-subject counts are reported in Table~\ref{tab:per-rat-counts}. Figure~\ref{fig:dataset-example}(a,b) shows a representative frame from animal Rat11, including the Jet-colormap thermal rendering and a colorized rendering of the mask used throughout this paper to display anatomical segmentations.
Table~\ref{tab:acq-summary} summarizes the acquisition parameters, and Table~\ref{tab:class-characteristics} reports the per-class composition and surface-temperature statistics computed over all $1{,}655$ frames. The four classes are strongly imbalanced: background occupies $86.9\%$ of all labeled pixels, the animal occupies only $13.1\%$ (median about $39{,}000$ pixels per frame), and the tail is the smallest class at $1.09\%$ (median about $3{,}200$ pixels per frame). This imbalance both motivates the inverse-frequency class weighting used in Section~\ref{sec-tech-validation} and explains the lower tail accuracy reported there. The per-class median temperatures follow a head~$>$~body~$>$~tail ordering, with the tail about $3.9\,^{\circ}$C cooler than the body; the ethanol cohort is warmer than the ketamine cohort across every class (Figure~\ref{fig:class-temp-dist}). All statistics in these tables and figures are reproducible from the raw data with the \texttt{dataset\_descriptors.py} script in the code repository.

\begin{table}[h]
    \centering
    \begin{tabular}{l l}
        \toprule
        \textbf{Property} & \textbf{Value} \\
        \midrule
        Thermal camera & FLIR ONE PRO (LWIR uncooled microbolometer) \\
        Sensor resolution & $640 \times 480$ (width $\times$ height) \\
        Subjects & 25 rats (15 ethanol, 10 ketamine) \\
        Imaging sessions & 41 \\
        Quality-controlled frames & $1{,}655$ \\
        Ambient temperature & $22$--$25\,^{\circ}$C \\
        \bottomrule
    \end{tabular}
    \caption{Acquisition summary for the released corpus, consolidating the parameters described in the Methods.}
    \label{tab:acq-summary}
\end{table}

\begin{table}[h]
    \centering
    \small
    \begin{tabular}{l
                    S[table-format=2.2]
                    S[table-format=6.0]
                    c
                    S[table-format=2.2]
                    S[table-format=2.2]}
        \toprule
        & & & \multicolumn{3}{c}{\textbf{Median temperature ($^{\circ}$C)}} \\
        \cmidrule(l){4-6}
        \textbf{Class} & {\textbf{Pixel \%}} & {\textbf{Med.\ px/frame}} & \textbf{Overall [IQR]} & {\textbf{Ethanol}} & {\textbf{Ketamine}} \\
        \midrule
        Background & 86.91 & 268253 & $22.07$ [$19.50$, $23.88$] & 23.36 & 21.75 \\
        Head       &  3.15 &   8463 & $28.35$ [$26.43$, $30.06$] & 29.11 & 28.25 \\
        Body       &  8.85 &  26078 & $28.07$ [$26.26$, $29.63$] & 28.66 & 27.97 \\
        Tail       &  1.09 &   3178 & $24.15$ [$22.36$, $25.96$] & 24.53 & 24.08 \\
        \midrule
        Animal (non-background) & 13.09 & 38947 & {--} & {--} & {--} \\
        \bottomrule
    \end{tabular}
    \caption{Per-class composition and surface temperature over all $1{,}655$ frames. \emph{Pixel \%} is the fraction of all labeled pixels and \emph{Med.\ px/frame} the median per-frame area on the $480 \times 640$ grid. Temperatures are medians: the \emph{Overall} column gives the median with the inter-quartile range $[Q_1, Q_3]$ in brackets, and the \emph{Ethanol} and \emph{Ketamine} columns are per-cohort medians. Across the corpus the per-frame median temperature ranges $15.3$--$28.7\,^{\circ}$C. The per-class medians follow a head~$>$~body~$>$~tail ordering, and the ethanol cohort is warmer throughout.}
    \label{tab:class-characteristics}
\end{table}

\begin{table}[h]
    \centering
    \small
    \begin{tabular}{l l S[table-format=4.0]}
        \toprule
        \textbf{Subject} & \textbf{Cohort} & {\textbf{Frames}} \\
        \midrule
        Rat1  & Ethanol  & 42 \\
        Rat2  & Ethanol  & 34 \\
        Rat3  & Ethanol  & 41 \\
        Rat4  & Ethanol  & 32 \\
        Rat5  & Ethanol  & 36 \\
        Rat6  & Ethanol  & 10 \\
        Rat7  & Ethanol  & 10 \\
        Rat8  & Ethanol  & 13 \\
        Rat9  & Ethanol  & 9 \\
        Rat10 & Ethanol  & 13 \\
        Rat11 & Ethanol  & 16 \\
        Rat12 & Ethanol  & 17 \\
        Rat13 & Ethanol  & 10 \\
        Rat14 & Ethanol  & 12 \\
        Rat15 & Ethanol  & 12 \\
        \midrule
        Ethanol subtotal & (15 subjects) & 307 \\
        \midrule
        Rat16 & Ketamine & 114 \\
        Rat17 & Ketamine & 120 \\
        Rat18 & Ketamine & 123 \\
        Rat19 & Ketamine & 130 \\
        Rat20 & Ketamine & 134 \\
        Rat21 & Ketamine & 133 \\
        Rat22 & Ketamine & 145 \\
        Rat23 & Ketamine & 140 \\
        Rat24 & Ketamine & 154 \\
        Rat25 & Ketamine & 155 \\
        \midrule
        Ketamine subtotal & (10 subjects) & 1348 \\
        \midrule
        \textbf{Total} & \textbf{(25 subjects)} & \textbf{1655} \\
        \bottomrule
    \end{tabular}
    \caption{Per-subject frame counts after QA. The ethanol cohort (Rats 1--15) yields relatively few frames per subject from short prevention-paradigm sessions, whereas the ketamine cohort (Rats 16--25) accumulates many more frames per subject across the longer therapeutic paradigm.}
    \label{tab:per-rat-counts}
\end{table}

\clearpage
\section{Technical Validation}\label{sec-tech-validation}
To illustrate that the dataset supports pixel-level anatomical analysis directly from the radiometric channel, we describe an exploratory U-Net semantic-segmentation pipeline that maps a temperature matrix to a four-class label map. The pipeline is reported here as a worked example of the kind of analysis the data afford, rather than as a methodological contribution in its own right. A qualitative output of the trained pipeline on Rat11 frame~1 is shown in Figure~\ref{fig:dataset-example}(c,d).

Each radiometric CSV 
yields the raw $480 \times 640$ temperature matrix in degrees Celsius. Because the cohort spans a wide temperature range, we adopt a fixed, physiologically motivated dynamic range: temperatures are clipped to the interval $[20, 40]\,^{\circ}$C and linearly rescaled to $[0,1]$, and the resulting single-channel image is replicated across three channels so that a standard ImageNet-pretrained encoder can be reused without architectural modification. To enforce a consistent geometry across acquisitions, both temperature matrices and masks are standardized to portrait orientation and resized to a $640 \times 480$ (height $\times$ width) grid with nearest-neighbor interpolation, which preserves the discrete class labels; any residual shape mismatch between a mask and its temperature matrix is likewise corrected by nearest-neighbor resampling. At training time the inputs are further augmented with horizontal flipping (probability $0.5$) and an affine perturbation (translation within $\pm 10\%$ of the image extent, isotropic scaling within $[0.9, 1.1]$, and rotation within $\pm 15^{\circ}$, applied with probability $0.5$, via the \texttt{Albumentations} library~\cite{buslaev2020albumentations}), while no augmentation is applied at validation or test time.

Generalization across subjects is assessed with a subject-level five-fold cross-validation in which folds are formed by a group $K$-fold scheme grouped on the rat identifier, so that all frames of a given animal fall in exactly one of the five test partitions. Within each outer fold, an inner group-shuffle split (random seed $42$, again grouped by rat) holds out one quarter of the non-test subjects as a validation set, ensuring that no animal contributes frames to more than one of the train, validation, and test partitions of a fold.

The network couples a U-Net decoder~\cite{ronneberger2015unet} with a ResNet-34 encoder~\cite{he2016deep} initialized from ImageNet pretraining, implemented through the \texttt{segmentation\_models\_pytorch} library~\cite{iakubovskii2019smp}, with three input channels matching the replicated thermal image and four output channels matching the class set. Training optimizes a combined objective with equal-weight ($0.5/0.5$) Dice~\cite{milletari2016vnet} and focal cross-entropy~\cite{lin2017focal} terms (focal focusing parameter $\gamma = 2.0$); to counter the imbalance between background and the smaller anatomical classes, per-class weights derived from the inverse square root of pixel frequency on the training corpus (cf.\ the class composition in Table~\ref{tab:class-characteristics}) are applied to both terms, taking values of approximately $0.22$ (background), $1.15$ (head), $0.69$ (body), and $1.94$ (tail). Optimization uses the Adam optimizer with a mini-batch size of $8$ in a two-stage transfer-learning schedule: the encoder is first frozen and only the decoder is trained at a learning rate of $10^{-3}$ for up to $20$ epochs, after which the full network is unfrozen and fine-tuned at $10^{-4}$ for up to $15$ further epochs; both stages use early stopping with a patience of $5$ epochs on the validation loss, and the checkpoint with the highest validation mean intersection-over-union is retained. At inference, the per-class logits are reduced to a discrete label map by pixel-wise argmax; because each frame contains a single animal, an $8$-connectivity connected-component analysis on the non-background prediction retains only the largest connected blob, suppressing spurious detections away from the main subject while leaving the internal head/body/tail partition untouched. The pipeline is implemented in Python using PyTorch together with the \texttt{segmentation\_models\_pytorch}, \texttt{Albumentations}, and OpenCV libraries, and runs on a CUDA-capable GPU when available, falling back to CPU otherwise.

\begin{figure}[!h]
    \centering
    \includegraphics{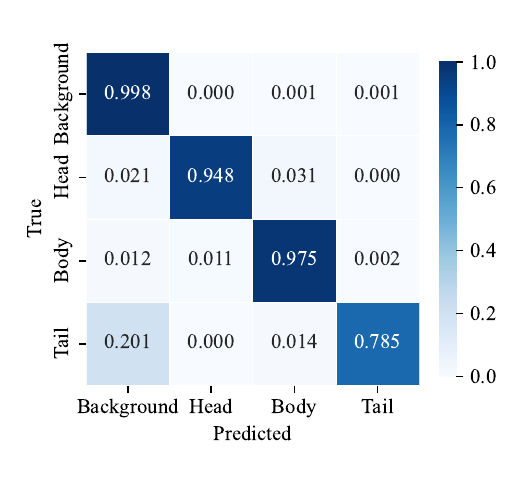}
    \caption{Aggregate test-set confusion matrix, summed across all five held-out partitions and row-normalized so that each row sums to $1$. Rows index the ground-truth class, columns the predicted class. The off-diagonal mass is dominated by Tail$\rightarrow$Body confusion, consistent with the lower per-class IoU reported for Tail in Table~\ref{tab:tv-kfold-results}.}
    \label{fig:tv-confusion}
\end{figure}

For each outer fold, the saved model is evaluated on its held-out subjects in terms of per-class intersection-over-union, the macro mean intersection-over-union across the four classes, and a $4 \times 4$ confusion matrix aggregating all test pixels. Table~\ref{tab:tv-kfold-results} summarizes the five-fold subject-level cross-validation over the released $1{,}655$-frame corpus. The pipeline attains a mean test mIoU of $0.895 \pm 0.006$ across folds. The Background, Body, and Head classes are recovered reliably (per-fold IoU above $0.99$, $0.95$, and $0.90$ respectively); the Tail class is the hardest, with a mean IoU of $0.724$ and an across-fold standard deviation of $0.025$, reflecting its small pixel footprint and frequent partial occlusion at the animal's distal end. Per-class means and medians agree to within $0.006$, indicating no strongly skewed-outlier behavior across folds. The aggregate confusion matrix in Figure~\ref{fig:tv-confusion}, summed over all five held-out partitions so that every animal contributes its pixels exactly once, confirms that the dominant residual error is Tail-to-Body confusion.

\begin{table}[!h]
    \centering
    \small
    \begin{tabular}{l c c c c c}
        \toprule
        \textbf{Fold} & \textbf{Test mIoU} & \textbf{BG} & \textbf{Head} & \textbf{Body} & \textbf{Tail} \\
        \midrule
        1 & 0.896 & 0.993 & 0.909 & 0.952 & 0.731 \\
        2 & 0.903 & 0.993 & 0.908 & 0.952 & 0.759 \\
        3 & 0.896 & 0.993 & 0.910 & 0.953 & 0.729 \\
        4 & 0.891 & 0.993 & 0.911 & 0.953 & 0.708 \\
        5 & 0.886 & 0.993 & 0.906 & 0.954 & 0.691 \\
        \midrule
        \textbf{Mean $\pm$ Std} & $0.895 \pm 0.006$ & $0.993 \pm 0.000$ & $0.909 \pm 0.002$ & $0.953 \pm 0.001$ & $0.724 \pm 0.025$ \\
        \textbf{Median} & $0.896$ & $0.993$ & $0.909$ & $0.953$ & $0.729$ \\
        \bottomrule
    \end{tabular}
    \caption{Per-fold test-set intersection-over-union for the 5-fold subject-level cross-validation. Each fold's row corresponds to the held-out subjects listed in the fold's \texttt{test\_rats.json}; per-class IoU and mIoU are computed from the confusion matrix accumulated over all test frames in that fold.}
    \label{tab:tv-kfold-results}
\end{table}

\section{Data Availability}
The dataset is openly available on Figshare at \href{https://doi.org/10.6084/m9.figshare.32309781}{\nolinkurl{10.6084/m9.figshare.32309781}}. It is organized as $25$ per-subject directories (\texttt{Rat1}--\texttt{Rat25}), each containing the parallel \texttt{CSV/}, \texttt{Mask/}, and \texttt{Thermal Imaging/} folders described in Section~\ref{sec:data-overview}, for a total of $1{,}655$ quality-controlled frames (approximately $0.45$\,GB compressed, $3.4$\,GB uncompressed). Each frame provides a raw $480 \times 640$ radiometric temperature matrix (CSV, degrees Celsius), a single-channel four-class \texttt{uint8} segmentation mask (PNG), and a Jet-colormap thermal rendering (PNG, or JPEG for a subset of animals). Loading conventions for all three artifacts are detailed in the Data Records section and in the accompanying \texttt{README}. The dataset is released under CC~BY~4.0.

\section{Code Availability}
The analysis code is openly available at \href{https://github.com/bykhov/rat_thermal_imaging_segmentaiton}{\nolinkurl{github.com/bykhov/rat_thermal_imaging_segmentaiton}}. The repository contains the complete 5-fold segmentation pipeline and a bundled sample with image examples.

\section*{Funding}
This research did not receive any specific grant from funding agencies in the public, commercial, or not-for-profit sectors.

\section*{Animal Ethics Statement}
All animal procedures were reviewed and approved by the Institutional Animal Care and Use Committee (IACUC) of Ben-Gurion University of the Negev and complied with the National Institutes of Health Guide for the Care and Use of
Laboratory Animals. The ketamine experiment was conducted under approval number \texttt{BGU-IL-2512-143-5}, and the ethanol experiment under approval number \texttt{BGU-358-08-2024-E}.

\bibliographystyle{ieeetr}
\bibliography{refs}





\end{document}